\documentclass[prl,aps,twocolumn,preprintnumbers,amsmath,amssymb,nofootinbib,superscriptaddress,notitlepage]{revtex4-1}
\usepackage[dvipdfmx]{graphicx}
\usepackage{ulem}
\usepackage[dvipsnames]{xcolor}
\usepackage{bm}
\usepackage{subfigure}
\usepackage{multirow}
\usepackage{amsmath}

\newcommand \cc[1]{{\bm {#1}}}
\newcommand \cb[1]{\overline{\bm {#1}}}
\newcommand \bQ{\bar Q}
\newcommand \drangle{\rangle\!\rangle}
\newcommand \dlangle{\langle\!\langle}
\newcommand \cstring{\!\rightarrow\!}

\begin{document}
%
%\setlength{\baselineskip}{16pt}
%
%---------------------------------------------------------------------------
%
\title{
Quark Confinement for Multi-Quark Systems \\
-- application to fully-charmed tetraquarks --
}
%
%----------------------------------
\author{
Guang-Juan Wang
}
\email[]{wgj@post.kek.jp}
\affiliation{KEK Theory Center, Institute of Particle and Nuclear Studies (IPNS), High Energy Accelerator
Research Organization (KEK), 1-1 Oho, Tsukuba, Ibaraki, 305-0801, Japan}
\affiliation{Advanced Science Research Center, Japan Atomic Energy Agency, 
Tokai, Ibaraki 319-1195, Japan}
\author{
Makoto Oka
}
\email[]{makoto.oka@riken.jp}%\thanks{corresponding author}%
\affiliation{Nishina Center for Accelerator-Based Science, RIKEN, Wako 351-0198, Japan}
\affiliation{Advanced Science Research Center, Japan Atomic Energy Agency, 
Tokai, Ibaraki 319-1195, Japan}

\author{
Daisuke Jido
}
\email[]{jido@th.phys.titech.ac.jp}
\affiliation{Department of Physics, Tokyo Institute of Technology, Meguro, Tokyo 152-8551, Japan
}
%
%--------------------------------
%
% @@ =======================================================================
%
\date{\today}

\begin{abstract}
A new color basis system and confinement mechanism for multi-quark systems are proposed according to 
the string-type picture of QCD.
The color string configurations in the strong coupling QCD are implemented in the set of color basis states.
The extended color Hilbert space for $QQ\bQ\bQ$ systems includes a ``hidden color'' state, which mixes 
with two-meson states $Q\bQ+Q\bQ$, This mixing effect leads to an attractive potential sufficient to form a bound state.
We apply a realistic Hamiltonian model with the new scheme to fully charmed tetraquark states, $cc\bar c\bar c$, and find
a bound and two resonant states, which could potentially correspond to the $cc\bar c\bar c$ tetraquark candidates recently observed in experiments.
\end{abstract}
\pacs{pacs numbers }
\maketitle
%
% @@ =======================================================================

Hadron spectroscopy in the past 70 years has established a rather simple view of hadrons  that
are classified into mesons made of a constituent quark ($Q$) and an anti-quark ($\bQ$), and baryons made of three quarks ($QQQ$).
While the majority of the observed hadrons fall into these categories, 
recent experiments have brought us a new kind of hadrons composed of more than three quarks \cite{Hosaka:2016pey,Chen:2016qju,Richard:2016eis,Olsen:2017bmm,Brambilla:2019esw,Liu:2019zoy,Guo:2017jvc,Meng:2022ozq}.
These exotic hadrons include tetraquarks ($QQ\bQ\bQ$), pentaquarks ($QQQQ\bQ$), and dibaryons ($QQQQQQ$).

Hadrons described by quantum chromodynamics (QCD), are strongly interacting systems of colored quarks and gluons with a non-trivial color confinement mechanism.
The string-type confinement potential is quite popular for mesons and baryons \cite{Wilson:1974sk,Kogut:1974ag,Carlson:1982xi,Isgur:1983wj,Isgur:1984bm}.
The string in the meson is simply a linear potential between $Q$ and $\bQ$.
For the baryon, the string configuration can be either a sum of two-body strings ($\Delta$-type),  
a three-body $Y$-type string  \cite{Takahashi:2000te, Takahashi:2002bw}, or their mixture \cite{Capstick:1986ter}, which are not well distinguished in the quark model
because they yield nearly identical mass spectra of low-lying states  \cite{Dmitrasinovic:2009dy,Dmitrasinovic:2009ma}.
On the other hand, in the multi-quark system, the strings may form richer topological configurations that do not appear in the $Q\bQ$ nor $QQQ$ hadrons~\cite{Jaffe:2007id,Dmitrasinovic:2001nu,Dmitrasinovic:2003cb,Green:1992yw,Okiharu:2005eg,Okiharu:2004wy}.

In studying the color confinement mechanism in multi-quark systems, the fully-heavy tetraquark, $QQ\bar Q\bar Q$ ($Q=c$, or $b$), is the most convenient and promising. The heavy quarks can be safely treated in the nonrelativistic Schr\"odinger equation. 
The $Q\bQ$ annihilation is suppressed due to the Okubo-Zweig-Iizuka (OZI) rule so that it can be omitted. 
The dynamics of the heavy quarks were studied very well in the quarkonium ($Q\bQ$) spectra, 
where the Hamiltonian parameters have been well-determined.

We here pay great attention to the recently observed fully charmed tetraquarks, $cc\bar c\bar c$ \cite{CMS:2016liw,CMS:2023owd,CMS:2020qwa,LHCb:2018uwm,LHCb:2020bwg,ATLAS:2023bft}.
Experimental data show a few resonant states. 
Comparison of these data with theoretical calculations will provide us with essential information about the confinement
mechanism of multi-quark systems.
There are indeed many theoretical studies done~\cite{Wang:2021kfv,Wang:2022yes,Albuquerque:2020hio,liu:2020eha,Jin:2020jfc,Lu:2020cns,Giron:2020wpx,Dosch:2020hqm,Yang:2020wkh,Huang:2020dci,Hughes:2021xei,Faustov:2021hjs,Liang:2021fzr,Li:2021ygk,Liu:2021rtn,Zhou:2022xpd,Asadi:2021ids,Liu:2021rtn,Yang:2021hrb,Ke:2021iyh,Wang:2021kfv,Guo:2020pvt,Dong:2020nwy,Gong:2022hgd,Wang:2020wrp,Gong:2020bmg,Guo:2020pvt,Liang:2022rew,Wang:2022jmb,Wan:2020fsk,Zhu:2020snb},
but no consensus has been reached on the structures of the observed resonant states. 

In the conventional quark model (QM), the color is carried only by  
the color $\cc3$ constituent quarks ($Q$) and the $\cb 3$ anti-quarks ($\bQ$).
Then the tetra-quark systems $QQ\bQ\bQ$ have two independent color-singlet states, 
and they are taken for example as the two combinations of $Q\bQ-Q\bQ$ states (QM basis) given by%
\footnote{Note that the choices of the two color configurations in Eq. \eqref{meson-basis} are not unique. Any two configurations of the tetraquark can be expressed in terms of each other, as can be proven by Fierz transformation.}
\begin{align}
\begin{aligned}
& |\cc1\rangle\equiv|(Q_1 \bQ_3)_{\cc1} (Q_2\bQ_4)_{\cc1}\rangle ,\\
& |\cc1'\rangle\equiv|(Q_1 \bQ_4)_{\cc1} (Q_2\bQ_3)_{\cc1}\rangle,
\label{meson-basis}
\end{aligned}
\end{align}
where $(Q \bQ)_{\cc1}$ denotes the color singlet state of $Q$ and $\bQ$, and the two states are independent but not orthogonal to each other as  $\langle \cc1'|\cc1\rangle=1/3$.
 The standard confinement potential is taken as the sum of the two-body color-dependent linear potentials \cite{Greenberg:1981xn}.\footnote{This confinement potential is known to lead to an unrealistic long-range color-van-der-Waals force~\cite{Miyazawa:1979vx,Greenberg:1979jw,Greenberg:1981xn,Isgur:1983wj,Isgur:1984bm,Lenz:1985jk,Oka:1984yx,Oka:1985vg,Koike:1986zx,Koike:1986mm,Koike:1986zp,Masutani:1987cb,Morimatsu:1989yn,Koike:1989ak,Horowitz:1991ux,Horowitz:1991fn,Koike:2000wk,Miller:1987wv,Vijande:2007ix,Vijande:2011im,Richard:2017vry,Deng:2017xlb,Martens:2006ac}.}
 In Ref.~\cite{Wang:2022yes}, they applied the conventional quark-model Hamiltonian determined well by the $Q\bQ$ systems to the tetra-quark systems with the complex scaling technique. Although they found a few resonant states in $0^{++}$ and $2^{++}$ channels, they come at much higher energies than recently found $cc{\bar c}{\bar c}$ candidate, $X(6900)$ \cite{CMS:2016liw,CMS:2020qwa,ATLAS:2023bft} and do not correspond to  $X(6600)$. We conjecture that the failure is due to defects of the conventional confinement mechanism when applied to multiquark systems.

\begin{figure}[thbp]
\centering\includegraphics[width=0.45\textwidth]{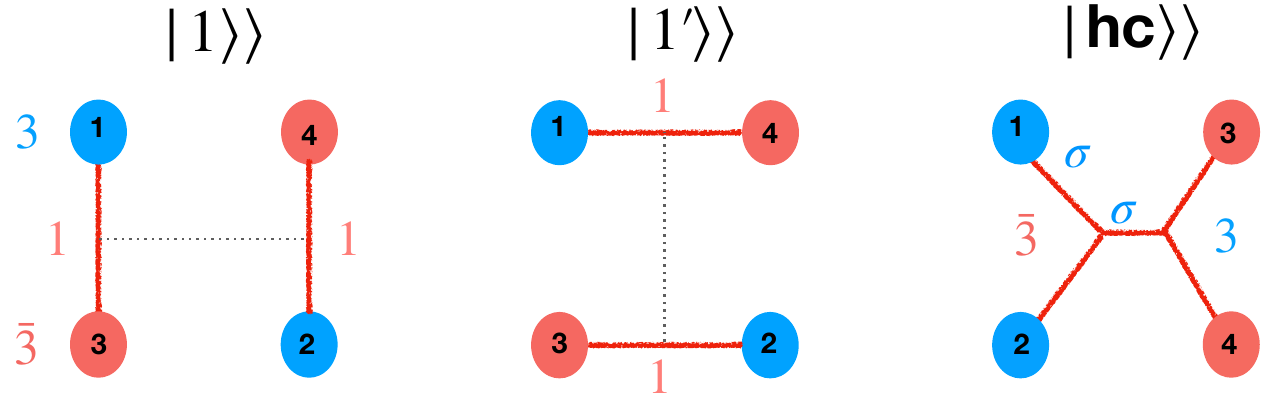}
\caption{Three color basis states: $| {\cc 1}\drangle$, $| {\cc 1'}\drangle$, and $|{\rm\bf hc}\drangle$  in the novel string-type color confinement model.}
\label{fig:tetraquark-color}
\end{figure}

Here, we propose an alternative string-type potential that represents the property of confinement of QCD. In the new confinement mechanism,  considering the topological properties of the color strings connecting $Q$ and $\bQ$ explicitly, we introduce a new set of color basis as shown in Fig. \ref{fig:tetraquark-color}. 

First of all, we introduce the meson-meson basis states by specifying the string connections as
\begin{eqnarray}
&& |\cc1\drangle\equiv|(Q_1\cstring\bQ_3)_{\cc1} (Q_2 \cstring\bQ_4)_{\cc1}\rangle , \\
&& |\cc1'\drangle\equiv|(Q_1\cstring\bQ_4)_{\cc1} (Q_2 \cstring\bQ_3)_{\cc1}\rangle,
\end{eqnarray}
where the arrow ($\cstring$) represents the direction of the color flux going from $Q$ to $\bQ$. As in the strong coupling limit of  QCD, these string-like states (ST basis) exhibit different topological structures that can be interchanged by the insertion of a vertex,  we impose the orthogonality \cite{Robson:1987pu,Miller:1987wv},
\begin{eqnarray}
&& \dlangle \cc1'|{\cc1}\drangle=0.
\end{eqnarray}
Thus the new ST basis is distinguishable from the QM basis in Eq. \eqref{meson-basis}.\footnote{A similar idea was discussed in the context of the string flip-flop potential model \cite{Morimatsu:1989yn,Koike:2000wk,Miller:1987wv,Alexandrou:1990zf,Pirner:1991im,Takahashi:2000te}.}

In addition,
we  introduce a new color basis state, $|{\rm\bf hc}\drangle$, named ``hidden color'' (HC), for a confined string configuration,
in which all four quarks are connected in one form of the strings (See Fig.~\ref{fig:tetraquark-color}).
Here we take a possible attractive configuration that $QQ$ and $\bQ\bQ$ form color $\cb{3}$ and $\cc{3}$, respectively. We suppose that
this configuration is independent and orthogonal both to $|{\cc 1}\drangle$ and $|{\cc 1'}\drangle$. 

In the current ST model, the three independent and orthogonal color configurations, $|{\cc 1}\drangle$, $|{\cc 1'}\drangle$ and $|{\rm\bf hc}\drangle$,  form the color Hilbert space.
The total wave function of the tetraquark system is given schematically by
\begin{eqnarray}
&& \Psi (1, 2, 3, 4) = \psi_{\bm 1}  | {\cc 1} \drangle + \psi_{{\bm 1}'}| {\cc 1'} \drangle
+ \psi_{\rm\bf hc} | {\rm \bf hc} \drangle,
\end{eqnarray}
where $\psi$'s are the orbital, spin, and flavor parts of the wave function.
The first two meson-meson (MM) states  describe asymptotic behaviors of meson-meson scattering states, 
while the HC term is confined in the range of color confinement and does not have asymptotic amplitude.  For the identical quarks, one needs to incorporate anti-symmetrization for the exchange of fermion in wave functions.

We introduce an explicit confinement potential for the new scheme of color basis. The new confinement potential is expected to properly describe the confined meson as well as the scattering state. The confinement potential, $V_{\rm conf}$, is given for three color basis states,
$|{\cc 1}\drangle$, $|{\cc 1'}\drangle$ and $|{\rm\bf hc}\drangle$,
by the following $3\times 3$ matrix,
\begin{widetext}
\begin{eqnarray} \label{eq:st}
&& V_{\rm conf} = 
\begin{pmatrix}
\sigma (r_{13}+r_{24}) & \kappa e^{-\sigma S}
& \kappa' e^{-\sigma S}\cr
\kappa e^{-\sigma S} & \sigma (r_{14}+r_{23}) &
 -\kappa' e^{-\sigma S}\cr
\kappa' e^{-\sigma S}&
-\kappa' e^{-\sigma S}&
\sigma [\frac14 (r_{13}+r_{24} +r_{14}+r_{23}) + \frac12 (r_{12}+r_{34})]
\cr
\end{pmatrix},
\end{eqnarray}
\end{widetext}
where $r_{ij}$ is the distance between the $(i,j)$ quark pair and $\sigma$ is the string tension.
The diagonal potentials for the meson and confined HC channel are chosen to go back to the conventional quark model due to its great success in the confined mesons and baryons.

The off-diagonal potentials describe the transitions among the color basis with different string topologies that can only happen with the insertion of a vertex due to the orthogonality. 
Motivated by strong coupling QCD,
 the string reconnection is induced by forming a color string surrounding the area,
like a Wilson loop.  The transition amplitude is given by the exponential of the minimal surface for the two 
reconnecting strings, $e^{-\sigma S}$,  where $S$ is the minimal surface of the tetraquarks \footnote{For computational simplicity, we approximate the minimal area $S$ by
\begin{eqnarray}
&& S= \frac{1}{4} ( r_{13}^2+r_{24}^2 + r_{14}^2 + r_{23}^2),
\end{eqnarray}
which agrees with the minimal surface 
for the square configuration, but, in general, overestimates the true value.}.
One sees that transitions may occur when all the quarks get together 
within the confinement region, which is characterized by  $1/\sqrt{\sigma} \le 0.5$ fm. In Eq.~(\ref{eq:st}), new coupling parameters $\kappa$ and $\kappa'$ are introduced.
As the transition of a string-anti-string pair to color octet state is $\sqrt{8}$ times larger than the color singlet state, 
we set $\kappa'=\sqrt{8}\kappa$ and vary $\kappa$ from 0.06 GeV to 0.14 GeV in the following calculation.

Now we apply the novel confinement potential model to the fully charmed tetraquark system $cc\bar c\bar c$.
The S-wave $cc\bar c\bar c$ systems exhibit three spin-parity combinations, 
$J^{PC}=0^{++}$, $1^{+-}$, and $2^{++}$ \footnote{The meson-meson scattering states involving higher orbital excited charmonium may modify the results of the higher $cc\bar c\bar c$ tetraquark states, which are beyond the scope of our investigation at this stage.}.

Our Hamiltonian for the $cc\bar c \bar c$ system is given by
\begin{eqnarray}
&& H = K +V_{\rm conf} + V_{\rm SR},\\
&& K=\sum_i \frac{{\bm p}^2_i}{2m_c}-{\frac{\bm P_{\text{tot}}^2}{8m_c}},\quad V_{\rm SR} = \sum_{i<j}  (T_i\cdot T_j) v_{ij}, \\
&& v_{ij}=  \frac{\alpha_s}{r_{ij}} -\frac{8\pi\alpha_s}{3m_c^2}\left(\frac{\Lambda}{\sqrt{\pi}}\right)^3e^{-\Lambda^2 r_{ij}^2} ({\bm s}_i\cdot{\bm s_j}), 
\end{eqnarray}
where $K$ is the kinetic energy with $\bm P_{\text{tot}}$ denoting the total momentum of the $cc\bar c\bar c$ state and $V_{\rm SR}$ represents the short-range interactions arising from the one gluon exchange
between two quarks with a two-body potential $v_{ij}$. The first  and second  terms in $v_{ij}$
correspond to the spin-independent color-Coulomb and the spin-dependent color-magnetic  interactions, respectively. 
The values of the parameters are given in Table \ref{tab:parameters}, which are fixed by the charmonium spectrum.

\begin{table}
\begin{tabular}{cccc}
\hline
$\sigma$ [GeV/fm]& $\alpha_s$& $m_c$ [GeV]& $\Lambda$ [GeV] \cr 
 \hline 0.7222 & 0.5461&  1.4794& 1.0946\cr
\hline
\end{tabular}
\caption{The parameters of the Hamiltonian.}\label{tab:parameters}
\end{table}%

We solve the bound and resonant states of the $cc{\bar c}{\bar c}$ system 
using the complex scaling method (CSM), which successfully reproduces the meson-meson scattering states with the correct positions of the thresholds. The technical details of CSM for this system are given in the
previous paper \cite{Wang:2022yes}. 
The detailed results will be provided in a separate paper.

\begin{figure}[htbp]
\centering\includegraphics[width=0.5\textwidth]{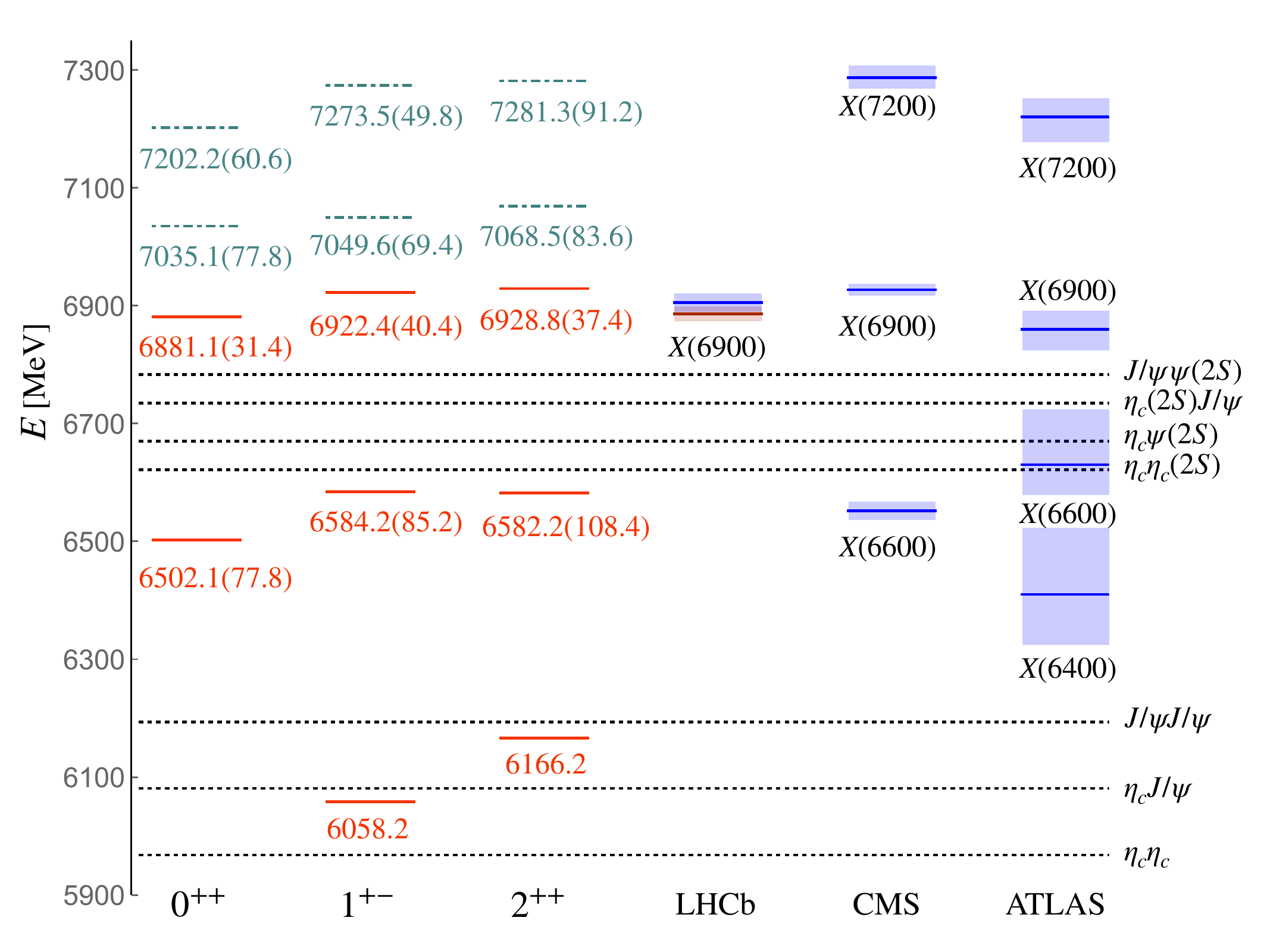}
\caption{Comparison of the $cc\bar{c}\bar{c}$ tetraquark spectrum using the novel string-type confinement mechanism with $\kappa=0.10$ GeV (red solid line), the conventional confinement potential (green solid-dot line) \cite{Wang:2022yes}, and the experimental data reported by LHCb \cite{LHCb:2020bwg}, CMS \cite{CMS:2023owd} (non-interference results),  ATLAS ($A$ and $\alpha$ fitting model) \cite{ATLAS:2023bft}, respectively. The theoretical results are presented by the mass $E$ and the decay width $\Gamma$ as $E$($\Gamma$) in units of MeV.}
\label{tetraquarks}
\end{figure}

Our numerical solutions for the overall $cc\bar c\bar c$ spectrum are summarized for a $\kappa$ value of $1.0$ GeV in Fig.~\ref{tetraquarks}
in comparison with experiments and the previous theoretical calculations~\cite{Wang:2022yes}. We successfully identify two resonant states with masses around $6.6$ GeV and $6.9$ GeV, which agree with the experimental observations $X(6600)$ and $X(6900)$, respectively. In the $1^{+-}$  and  $2^{++}$ sectors, the results for resonances are similar, which is consistent with the heavy quark spin symmetry.  Additionally, we observe a hint of a higher state in the complex plane 
in spite of bad convergence in the CSM, which is not shown in the figure. This state needs to be checked in the future, for which the contributions from scattering states involving the higher orbital excitations may be significant.   On the experimental side, 
a higher resonance is suggested as  $X(7200)$ or $X(7000)$ in CMS \cite{CMS:2023owd}  and ATLAS ($\beta$ fitting model)\cite{ATLAS:2023bft}, respectively. Further experimental investigations, such as the resonant position and identification of their quantum numbers, are required to study the properties of this resonance. 

 Our results show the existence of the two bound states. Their binding energies are similar to each other and the mass difference arises from the difference of the scattering thresholds, $\eta_c$-$J/\psi$, and $J/\psi$-$J/\psi$, respectively. The existence of the bound state below the $J/\psi$-$J/\psi$ threshold should be examined in future experiments.
On the other hand, no $0^{++}$ bound state is found, the reason for which will be discussed later.

\begin{figure}[htbp]
\centering\includegraphics[width=0.5\textwidth]{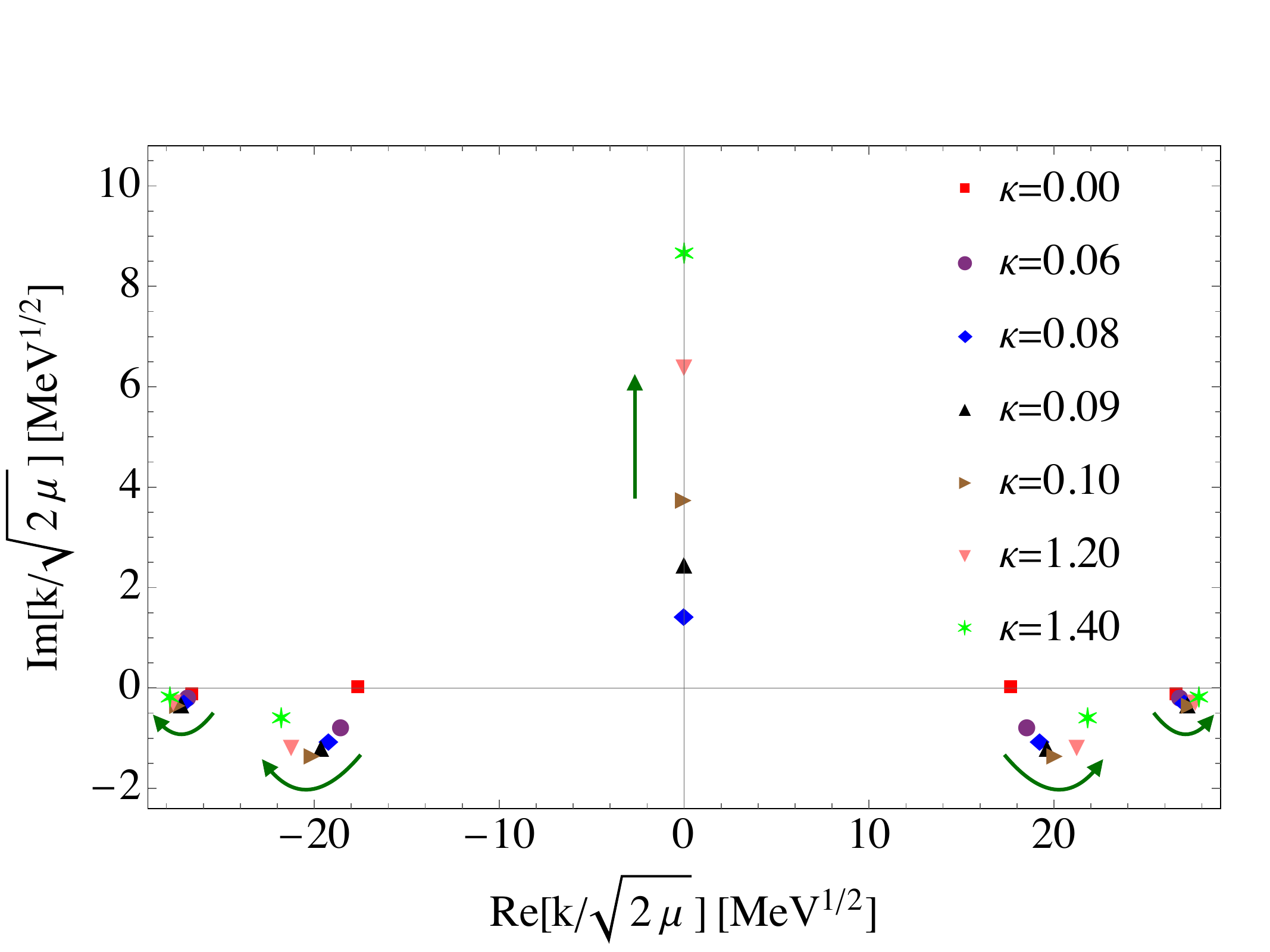}
\caption{The trajectories of the bound and resonant states with $J^{PC}=2^{++}$  plotted on the complex energy plane 
for varied $\kappa$ from $0.00$ GeV to $0.14$ GeV. The direction of variation with increasing $\kappa$ is indicated by arrows.
The origin of the complex plane is set to the $J/\psi$-$J/\psi$ threshold and $\mu$ is the reduced mass. 
}
\label{poles}
\end{figure}

To understand the dynamical origin of the bound state and compare the novel confinement model with the conventional one, we focus on examining the spin-aligned S-wave ($L=0$) $cc\bar c\bar c$ state with $ J^{PC}= 2^{++}$ and evaluate the impact of the chosen coupling constant $\kappa$.
We summarized the obtained $cc\bar c \bar c$ states with $\kappa$ varying from $0$  to $0.14$~GeV  in  Table~\ref{tab: 2++12} and plot the  ``motion'' of the states in the complex momentum plane in Fig.~\ref{poles}. 

Our results show a bound state for $\kappa$ larger than 0.08~GeV, and the binding energy increases progressively as $\kappa$ increases\footnote{It should be noted here that a virtual state may appear for $\kappa$ ranging from $0.06$ and $0.80$ GeV at the negative imaginary axis, but
the virtual state is not an eigenvalue of the present complex scaling Hamiltonian.}. 
The probabilities of the meson-meson state, $P$(MM), and the hidden color state, $P$(HC),
shown in Table~\ref{tab: 2++12}, exhibit molecular characteristics of the bound states. 
$P$(HC) increases as the binding energy grows.
These behaviors show that the mixing of the HC configurations is the driving
force for the formation of the bound state. Notably, in the previous study~\cite{Wang:2022yes}, the same short-range $V_{SR}$ but with the conventional confinement potential were used and they obtained no bound states. Furthermore, the resonances appeared at higher masses than the observed states. The reason is that, in the conventional scheme, the hidden color state is not independent of the mesonic states.  The mixing effect was not as pronounced as it is in the new confinement potential, resulting in a weaker binding effect and no appearance of bound states. This helps explain the absence of the bound tetraquark state in $0^{++}$ sector with $\kappa=0.1$ GeV, where the hidden color (HC) component will couple with two meson-meson channels,  $\eta_c$-$\eta_c$, and $J/\psi$-$J/\psi$. Compared with the $2^{++}$ sector, the mixing effects arising from $\kappa e^{-\sigma S}$ in $0^{++}$ are suppressed by the extra spin overlap factors, $\frac{1}{2}$ for $\eta_c$-$\eta_c$ and $\frac{\sqrt{3}}{2}$ for $J/\psi$-$J/\psi$, leading to smaller attractive potentials in the meson-meson channels. 

For the resonances in Fig.~\ref{poles}, as $\kappa$ increases, the resonant states shift towards higher masses with larger decay widths and then exhibit a bending behavior, returning toward the real axis. The decreasing widths of the dressed tetraquarks as the coupling increase are consistent with the curved trajectories discussed in Refs.~\cite{Hanhart:2022qxq,vanBeveren:2006ua,Ortega:2021fem}, which are attributed to the strong unitarization effect.

\begin{table}
  \centering
\setlength{\tabcolsep}{2.3mm}
\begin{tabular}{c|cc|c|c}
\hline 
 & \multicolumn{2}{|c|}{Bound state} &1st Resonance &2nd Resonance\tabularnewline
\hline 
\multirow{2}{*}{$\kappa$} &  $E$ & $\Delta E$ & \multirow{2}{*}{$E-  \frac{\Gamma}{2} i$} 
& \multirow{2}{*}{$E-  \frac{\Gamma}{2} i$} \tabularnewline
&  $P$(MM)  & $P$(HC) &  &  \tabularnewline
\hline
$0.06$ &  - &  & {$6524.4-28.6i$}  &{$6900.8-9.5i$} \tabularnewline
\hline 
\multirow{2}{*}{$0.08$} & $6179.5$  & $-2$ & \multirow{2}{*}{$6550.2-41.0i$} & \multirow{2}{*}{$6912.7-16.6i$}\tabularnewline
 &  $98.2\%$& $1.8\%$   & \tabularnewline
\hline 
\multirow{2}{*}{$0.09$} &  $6175.0$ & $-6$ & \multirow{2}{*}{$6564.9-47.3i$} 
& \multirow{2}{*}{$6920.8-18.3i$}\tabularnewline
&  $93.1\%$  & $6.9\%$   & \tabularnewline
\hline 
\multirow{2}{*}{$0.10$} &  $6166.2$ &  $-14$ & \multirow{2}{*}{$6582.2-54.2i$} 
&  \multirow{2}{*}{$6928.8-18.7i$}  \tabularnewline
& $88.3\%$ & $11.7\%$   & \tabularnewline
\hline 
\multirow{2}{*}{$0.12$} & $6139.8$ & $-41$ & \multirow{2}{*}{$6630.6-50.6i$} 
 & \multirow{2}{*}{$6943.7-16.3i$} \tabularnewline
 &  $80.6\%$ & $19.4\%$  & \tabularnewline
\hline 
\multirow{2}{*}{$0.14$} & $6106.1$ & $-74$  &\multirow{2}{*}{$6655.5-27.5i$} 
 & \multirow{2}{*}{$6955.0-11.8i$}\tabularnewline
&  $75.2\%$& $24.8\%$  &\tabularnewline
\hline 
\end{tabular}
\caption{The obtained states for $J^{PC}=2^{++}$ with $\kappa$ varying from $0.06$ GeV to $0.14$ GeV: the real part $E$ and the imaginary part $\frac{\Gamma}{2}$ (in units of MeV)
 of the bound state and two low-lying resonances.
$P$(MM) and $P$(HC) are the probability of the meson-meson and hidden color states for the bound state, respectively.}
\label{tab: 2++12}
\end{table}

\bigskip
In conclusion, we have presented a new string-type model for the color Hilbert space and color confinement potential in multi-quark systems. The key intention is to extend the color Hilbert space with two channels of color singlet mesons and another hidden-color compact state. The third state is not allowed in the conventional quark model, where only the quarks carry the color degrees of freedom.  It is found that the couplings between the meson-meson (MM) channels to the hidden color (HC) channel are significant and induce a strong attraction.

We have developed a realistic Hamiltonian model for the S-wave $cc\bar c\bar c$ tetraquark system,  taking into account the 
color-Coulomb and color-magnetic interactions. The tetraquark spectrum is obtained by solving
the four-body problem with the complex scaling method. A bound state
below the $J/\psi$-$J/\psi$ threshold is found for a moderate coupling strength among the color channels,  which is not predicted by the conventional quark model. The bound states serve as a valuable benchmark to investigate the two different confinement mechanisms. Additionally,
we have identified two resonant states, which may correspond to the observed $cc\bar c\bar c$ resonant
states $X(6600)$ and $X(6900)$.
By determining the spin and parity of the observed states, we will be able to fix the coupling
parameter and draw a complete picture of the fully charmed tetraquark states.

%
% @@@ =================================================================
%
%\section*{Acknowledgments}
\bigbreak
We acknowledge Drs.~Toshiki Maruyama, Osamu Morimatsu, Atsushi Hosaka, Masayasu Harada, and Emiko Hiyama for useful discussions.
This work is supported by Grants-in-Aid for Scientific Research No.~JP20F20026, JP20K03959, JP21H00132, JP21K03530, JP22H04917, JP23K03427.

\bibliography{cccc.bib}

\end{document}